\documentclass[11pt, a4paper]{article}
\usepackage[english]{babel}
\usepackage{cite}
\usepackage{leftidx}
\usepackage[letterpaper,top=2cm,bottom=2cm,left=3cm,right=3cm,marginparwidth=1.75cm]{geometry}

\usepackage{amsmath,amssymb,amsfonts}
\usepackage{subfigure}
\usepackage{graphicx}
\usepackage[colorlinks=true, allcolors=blue]{hyperref}
\usepackage{cleveref}
\usepackage[page]{appendix}
\usepackage{algorithm}
\usepackage{algorithmic}
\usepackage{url}
\usepackage{textcomp}
\usepackage[affil-it]{authblk}
\usepackage[font=small,labelfont=bf]{caption}

\font\myfont=cmr12 at 22pt
\title{{\myfont On nonlinear compression costs: when Shannon meets R\'enyi}}
\author[1,2]{Andrea Somazzi\thanks{Corresponding author: {\href{mailto:andrea.somazzi@imtlucca.it}{andrea.somazzi@imtlucca.it}}}}
\author[3]{Paolo Ferragina\thanks{{\href{mailto:paolo.ferragina@unipi.it}{paolo.ferragina@unipi.it}}}}
\author[1,4,5]{Diego Garlaschelli}
\affil[1]{\footnotesize IMT School for Advanced Studies, Piazza S. Francesco 19, 55100 Lucca, Italy}
\affil[2]{Scuola Normale Superiore, Piazza dei Cavalieri 7, 56126 Pisa, Italy}
\affil[3]{Department of Computer Science, University of Pisa, Pisa, Italy}
\affil[4]{Lorentz Institute for Theoretical Physics, Niels Bohrweg 2, 2333 CA Leiden, The Netherlands}
\affil[5]{INdAM-GNAMPA Istituto Nazionale di Alta Matematica, Italy}

\date{}

\begin{document}
\maketitle
\begin{abstract}
Shannon entropy is the shortest average codeword length a lossless compressor can achieve by encoding i.i.d. symbols. However, there are cases in which the objective is to minimize the \textit{exponential} average codeword length, i.e. when the cost of encoding/decoding scales exponentially with the length of codewords. The optimum is reached by all strategies that map each symbol $x_i$ generated with probability $p_i$ into a codeword of length $\ell^{(q)}_D(i)=-\log_D\frac{p_i^q}{\sum_{j=1}^Np_j^q}$. This leads to the minimum exponential average codeword length, which equals the R\'enyi, rather than Shannon, entropy of the source distribution. We generalize the established Arithmetic Coding (AC) compressor to this framework. We analytically show that our generalized algorithm provides an exponential average length which is arbitrarily close to the R\'enyi entropy, if the symbols to encode are i.i.d.. We then apply our algorithm to both simulated (i.i.d. generated) and real (a piece of Wikipedia text) datasets. While, as expected, we find that the application to i.i.d. data confirms our analytical results, we also find that, when applied to the real dataset (composed by highly correlated symbols), our algorithm is still able to significantly reduce the exponential average codeword length with respect to the classical `Shannonian' one. Moreover, we provide another justification of the use of the exponential average: namely, we show that by minimizing the exponential average length it is possible to minimize the probability that codewords exceed a certain threshold length. This relation relies on the connection between the exponential average and the cumulant generating function of the source distribution, which is in turn related to the probability of large deviations. We test and confirm our results again on both simulated and real datasets.
\end{abstract}

\section{Introduction}
\label{sec:introduction}
In the realm of (lossless) data compression, the main goal is to efficiently represent data in a manner that requires reduced space without compromising its integrity. At the heart of this challenge lies the encoding strategy, which determines how individual symbols or sequences of symbols are transformed into compressed representations. Traditionally, these strategies aim to minimize the average length of the encoded symbols. By achieving a shorter average encoded symbol length, one can ensure a more compact representation of the entire input data, thereby achieving the central objective of many data compression problems. 

Consider a stationary source generating symbols from an alphabet $\Sigma=\{x_1,\dots,x_N\}$ of size $|\Sigma|=N$, with probability $p=\{p_1,\dots,p_N\}$. Then, the problem consists in finding the encoding strategy which maps each symbol $x_i \in \Sigma$ into a $D$-ary codeword of length $\ell_D(i)$ such that 
\begin{align}
    L(0) = \sum_{i=1}^N p_i \ell_D(i)
    \label{eq:lincost}
\end{align} is minimized. $L(0)$ is the codewords' average length, and the use of such notation will be clarified later. 

In his pioneering work~\cite{shannon1948mathematical}, Shannon proved that for a source generating i.i.d. symbols, Eqn.~\eqref{eq:lincost} is minimized by all encoding strategies such that $\ell_D(i)=-\log_D p_i$, for all $i = 1, 2, \ldots, N$. 
However, in most cases, strategies that guarantee such equality for each symbol do not exist but only get `close' to it. This leads to the notorious relation \begin{align}
    L(0) \geq H_1[p],
    \label{eq:boundsh}
\end{align} where $H_1[p]=-\sum_{i=1}^N p_i\, \log_D p_i$ is the Shannon entropy of the source, which can be understood as the codewords' minimum average length. The use of the subscript in $H_1$ will also be clarified later. 

We would also like to mention that Eqn.~\eqref{eq:lincost} can be seen as a \textit{cost function} $C$, because minimizing Eqn.~\eqref{eq:lincost} is equivalent to minimizing the cost of encoding/decoding $C(0) \propto L(0)$ under the assumption that such cost is linear in the codewords' length.

Beyond the conventional focus on the linear average of codeword lengths, it's essential to acknowledge that this is not the only viable metric to target for minimization. For example, there could be a \emph{nonlinear relation} between the cost of encoding/decoding symbols and their codewords' length. Delving deeper into the theoretical underpinnings of averages, we encounter the Kolmogorov-Nagumo (KN) averages~\cite{kolmogorov1930notion, nagumo1930klasse}: a more general family of averages that offers a richer landscape for exploration. One might be driven to consider minimizing these KN averages, recognizing the possibility of uncovering novel compression strategies and further refining data representation techniques that are suitable in different scenarios. Following the introduced notation, the codewords' KN average length is defined as \begin{align}
    \langle \ell_D \rangle_{\varphi} = \varphi ^{-1} \bigg (\sum_{i=1}^N p_i\;  \varphi (\ell_D(i)) \bigg),
    \label{eq:KNlen}
\end{align}
where $\varphi$ is a continuous injective function. Note that for $\varphi(x) = x$ the usual average length ~\eqref{eq:lincost} is recovered. While, in general, KN averages depend on $\varphi$, there is a natural requirement that an average length measure should satisfy, that restricts the space of admissible functions~\cite{aczel1975measures, campbell1966definition}. Namely, it should be \emph{additive} for independent symbols. In particular, consider two independent sets of symbols $\Sigma^{(1)}=\{x_1,\dots,x_N\}$ and $\Sigma^{(2)}=\{y_1,\dots,y_M\}$, respectively. The associated probabilities are $p=\{p_1,\dots,p_N\}$ and $q=\{q_1,\dots,q_M\}$, and each symbol is encoded in a codeword of length $\{\ell_D^{(1)}(i)\}_{i=1}^N$ and $\{\ell_D^{(2)}(j)\}_{j=1}^M$. Then, the additivity requirement is formulated as follows: \begin{align}
\begin{split} 
    &\varphi ^{-1} \bigg (\sum_{i=1}^N \sum_{j=1}^M p_i q_j\: \varphi \big(\ell_D^{(1)}(i)+ \ell_D^{(2)}(j)\big) \bigg)\\
    &=\varphi ^{-1} \bigg (\sum_{i=1}^N \: p_i \varphi (\ell_D^{(1)}(i)) \bigg) + \varphi ^{-1} \bigg (\sum_{j=1}^M q_j  \varphi (\ell_D^{(2)}(j)) \bigg).
    \label{eq:additivity}
    \end{split}
\end{align}
It is possible to prove that Eqn.~\eqref{eq:additivity} leads to the so-called exponential KN averages~\cite{ campbell1966definition,hardy1952inequalities, Morales_2023}, that correspond to $\varphi(x)=\varphi_t(x) = \gamma D^{tx}+b$. Substituting $\varphi_t$ into Eqn.~\eqref{eq:KNlen}, one gets that \begin{align}
    \langle \ell_D \rangle_{\varphi_t} \equiv L(t) = \frac{1}{t}\log_D\bigg(\sum_{i=1}^N p_i\:  D^{t\, \ell_D(i)}\bigg),
    \label{eq:expcost}
\end{align}
where $t>-1$ and $L(t)$ is then the \emph{exponential average} of the codeword's length. Notice that for $t$ approaching $0$, the exponential average converges to the linear average, i.e. $\lim_{t \to 0}L(t)=L(0)$, which clarifies the notation we have adopted before.

The utility of the exponential average in data compression can be understood from two distinct fronts. Firstly, when the costs associated with the encoding or decoding steps amplify ($t>0$), they might grow in an exponential fashion with respect to the codewords' lengths. This leads to a nonlinear relation having the form $C(t) \propto \sum_i p_i D^{t\ell_D(i)}$. Minimizing $C(t)$ is then equivalent to minimizing Eqn.~\eqref{eq:expcost} if $t>0$, since the latter is a monotonically increasing function of the former. A case falling in such scenario could be DNA coding~\cite{meiser2022synthetic, mishra2020compressed}, where the apparatus involved in encoding and decoding procedures is very costly. Minimizing this exponential cost function then could become essential for effective and efficient data handling.
Secondly, at a more theoretical level, the exponential average arises naturally when aiming at curtailing the risk of buffer overflow~\cite{buffer1, buffer2} or bolstering the probability of transmitting a message in a short timeframe. This could be the case in aerospace communication scenarios, where it can happen that antennas are visible for fleeting moments, necessitating the rapid and reliable transmission of information~\cite{iridium}.
In such scenarios, estimating the likelihood of large deviations (for the events to be avoided) involves the cumulant generating function of the probability distribution, which in turn leads to the exponential average.
Finally, it has been shown that minimizing Eqn.~\eqref{eq:expcost} with $t<0$ is a problem related to maximizing the chance of receiving a message in a single snapshot~\cite{baer2008optimal}.

In his valuable paper~\cite{campbell1965coding}, Campbell proved that the optimal encoding lengths that minimize the exponential cost of Eqn.~\eqref{eq:expcost} are \begin{align}
    \ell_D^{(q)}(i)=-\log_D\frac{p_i^q}{\sum_{j=1}^Np_j^q},
    \label{eq:escortlen}
\end{align}
where $q=1/(1+t)$. Moreover, he proved that the lower bound for the exponential cost is given by the R\'enyi entropy of order $q=1/(1+t)$ of the source, defined as \begin{align}
    H_q[p]=\frac{1}{1-q}\log_D\bigg(\sum_{i=1}^N p_i^q\bigg),
    \label{eq:renyi}
\end{align}
so that \begin{align}
    L(t) \geq H_{\frac{1}{1+t}}[p]
    \label{eq:boundren}
\end{align} where the equality holds iff Eqn.~\eqref{eq:escortlen} is exactly satisfied. Note that $\lim_{q\to 1}H_q[p]=H_1[p]$, i.e. Shannon entropy is a particular case of R\'enyi entropy. It follows that for $t \to 0$, Eqn.~\eqref{eq:boundren} reduces to Eqn.~\eqref{eq:boundsh}.

The probability distribution $p^{(q)}=\bigg\{\frac{p_1^q}{\sum_{j=1}^Np_j^q},\dots,\frac{p_N^q}{\sum_{j=1}^Np_j^q}\bigg\}$ which appears in Eqn.~\eqref{eq:escortlen} is often referred as \textit{escort} or \textit{zooming} probability distribution of $p$~\cite{tsallis2009introduction, beck_schögl_1993, somazzi2023learn}. The reason is that, depending on the value of $q$, it can amplify/suppress values in the tails of the original distribution $p$ (and, since it is normalized, suppress/amplify the others). Escort distributions have been applied and have emerged in various fields, ranging from non-extensive statistical mechanics~\cite{tsallis2009introduction}, chaotic systems~\cite{beck_schögl_1993} and statistical inference~\cite{somazzi2023learn}. 
Another notable link among the R\'enyi entropy, the KN exponential average, and escort distributions comes from an axiomatic point of view. While Shannon entropy can be derived by the four Shannon-Khinchin axioms (SK1-SK4)~\cite{khincin}, R\'enyi entropy is derived by relaxing SK4 (also called \textit{additivity} axiom) to a more general version, which involves both the KN exponential average and the escort distributions~\cite{jizba2020shannon}.

Since Campbell, from the point of view of data compression problems, escort distributions are also the optimal distributions according to which one has to encode symbols in order to minimize the exponential average codeword length $L(t)$. However, although Campbell provided the existence of an optimal encoding length, he did not suggest any operational strategy to achieve it. Some specific algorithms have been later proposed~\cite{buffer1, buffer2, merhav1991universal, blumer1988renyi}, and~\cite{bercher2009source} noted that, since the optimal lengths defined in Eqn.~\eqref{eq:escortlen} have the same form of the lengths which minimize the linear average length of Eqn.~\eqref{eq:lincost} if $p$ is replaced by its escort $p^{(q)}$, then it is sufficient to feed a standard (i.e. `Shannonian') encoder with $p^{(q)}$ instead of $p$ in order to reach a cost $L(t)$ close to its minimum $H_{\frac{1}{1+t}}[p]$.

In this paper, we provide a series of contributions. i) We lay the mathematical ground to the observations of the previous papers by applying the above conceptual framework to one of the most efficacious algorithms in the realm of data compression: i.e.,  Arithmetic Coding (AC) (Sec.~\ref{sec:methods}). ii) We experimentally analyze the performance of the proposed escort distribution-based compressor in the case of optimizing the exponential average codeword length, over both synthetic and real datasets. We confirm the theoretical results on the former (composed by i.i.d. generated symbols) and achieve surprising results on the latter (composed by correlated symbols). In particular, we show that on a sample of Wikipedia text the application of our compressor with escort probability leads to an improved compression ratio (when the considered metric is the exponential average codeword length) with respect to a standard Shannon compressor, even if the optimal value of $q$ (i.e. the exponent leading to the escort distribution) is unknown to the encoder (Sec.~\ref{sec:wikipedia}). iii) Finally, we examine analytically and experimentally the practical case in which its crucial to not exceed a certain threshold in the codewords' lengths (such as in the context of bounded buffers), by showing that the exponential average naturally appears in the probability of large deviations thus further justifying the study performed in the present paper. In particular, we will show that by using our approach it is possible to significantly reduce the probability that the length of the codeword assigned to a given sequence of symbols exceeds a certain threshold with respect to a classic Shannon compressor (Sec.~\ref{sec:discussion}).

It goes without saying that all our results and experimental achievements could benefit of the use of more recent statistical compressors (i.e., ANS \cite{moffat2020large}) in place of the arithmetic coder, whose simplicity is exploited in this paper just for clarity of explanation.

\section{Methods}
\label{sec:methods}
In the ensuing section, we undertake an examination of the arithmetic coding compression scheme. We commence by providing a theoretical description of AC, delineating its operating principles. Following this, we weigh the pros and cons of AC, offering a balanced viewpoint on its utility and limitations in various application contexts. Finally, we advance the discourse by generalizing AC with an aim to achieve the theoretical limit as predicted by Campbell's theorem.
\subsection{Arithmetic coding}
\label{sec:ac}
Arithmetic coding is a lossless encoding scheme~\cite{witten1987arithmetic}. Compressor and decompressor both need the alphabet of symbols $\Sigma$, the associated probability distribution $p$ and the length of the stream of symbols to encode/decode. Consider a string $\vec{s}=(s_1,\dots,s_M)$ of length $M$, where each $s_j=x_{i_j}$ is a symbol randomly generated by a source from alphabet $\Sigma$ with associated probability $p$. Then, in order to encode $\vec{s}$ into a $D$-ary alphabet, the encoder performs the procedure illustrated in Algorithm~\ref{alg:ac}. 

\begin{algorithm}[ht]
\caption{Arithmetic Coding}\label{alg:ac}
\begin{algorithmic}[1]
\REQUIRE The input string $\vec{s} = x_{i_1} x_{i_2} \cdots x_{i_M}$, the probabilities $p=\{p_1,\dots,p_N\}$ and the cumulative $f=\{f_1,\dots,f_N\}$ of $p$.
\ENSURE A subinterval $[a, a+\mathcal{S})$ of $[0,1)$.
\STATE $\mathcal{S}_0=1$
\STATE $a_0=0$
\STATE $j=1$
\WHILE{$j\neq M$}
\STATE $\mathcal{S}_j = \mathcal{S}_{j-1} \cdot p_{i_j}$
\STATE $a_j = a_{j-1} + \mathcal{S}_{j-1}\cdot f_{i_j}$
\STATE $j=j+1$
\ENDWHILE
\STATE {\bf return}\ $\langle k \in [a_M, a_M+\mathcal{S}_M), M\rangle$
\end{algorithmic}  
\end{algorithm}
Essentially the encoder, starting from the interval $[0,1)$, iteratively divides it proportionally to the probabilities in $p$ and, at each iteration $j$, chooses the subinterval corresponding to the associated symbol $s_j=x_{i_j}$. After $M$ iterations, the encoder emits a number $k$, contained in the final subinterval $[a_M, a_M+\mathcal{S}_M)$, with $\mathcal{S}_M=\prod_{j=1}^M p_{i_j}$, which is uniquely associated with the original string $\vec{s}$. Such number $k$ is then converted into its $D$-ary representation and communicated to the decoder (together with the original string length $M$), which can reverse this procedure to get the original string. It follows that the encoded string's length $\ell_D(\vec{s})$ is equal to the number of symbols (bits if $D=2$) necessary to encode $k$ in the desired alphabet. 

From now on, we will consider for simplicity a binary ($D=2$) encoding alphabet. Nonetheless, while our focus is on the classic binary AC, the results we present are inherently generalizable to $D>2$, ensuring that the core features and principles of AC we discuss remain applicable and valid to those other cases too.

It is possible to show that the length of the encoded number $k$ (i.e. encoded string) depends only on the length of the final subinterval $\mathcal{S}_M$. In particular, by choosing $k=a_M + \mathcal{S}_M/2$ and by truncating its binary representation to the first $\lceil \log_2\frac{2}{\mathcal{S}_M}\rceil$ bits, the approximation error is so small that such truncation is guaranteed to fall into the interval $[a_M, a_M+\mathcal{S}_M)$. Considering that \begin{align}
    \ell_2(\vec{s})=\bigg\lceil \log_2\frac{2}{\mathcal{S}_M}\bigg\rceil < 2-\log_2\mathcal{S}_M = 2-\sum_{j=1}^M \log_2 p_{i_j},
\end{align}
and that it is possible, for $M$ large enough, to approximate each $p_i$ with the fraction of occurrences of symbol $x_i$ in $\vec{s}$, i.e $p_i\simeq n_i(\vec{s})/M$, one gets that: \begin{align}
    \ell_2(\vec{s}) < 2+M\cdot H_0[p].
    \label{eq:acshbound}
\end{align}
Eqn.~\eqref{eq:acshbound} unveils the main strength of the AC scheme: the number of bits that are `wasted' in encoding $\vec{s}$ is $2$, thus resulting \emph{intensive} with respect to the string length $M$ (provided that we operate with infinite precision arithmetic~\cite{witten1987arithmetic}). As $M$ increases, the number of wasted bits per character goes to $0$, in fact \begin{align}
    \frac{\ell_2(\vec{s})}{M} < \frac{2}{M} \cdot H_0[p].
\end{align}
A primary limitation of arithmetic coding (AC) lies in its operational framework. Unlike certain encoding schemes that allocate distinct codewords to individual symbols, AC assigns a codeword to the entire string. This means that the decoding process cannot commence in tandem with encoding so the decoder must wait for the encoder's completion of encoding the entire string  (see e.g. the variant Range Coding for relaxing this limitation~\cite{ferragina_2023}). As the efficiency of AC generally improves with an increase in $M$, this waiting period can be time-consuming, rendering AC unsuitable for some applications. Conversely, AC boasts superior performance compared to encoding mechanisms that designate codewords to each symbol particularly when probability distributions are highly skewed. Such encoders mandate a minimum of 1 bit per symbol. However, the optimal length — expressed as $-\log_2p_i$ — can be significantly less than 1.

\subsection{Generalized AC}
\label{sec:acq}
We now propose a generalization of AC in order to optimally minimize the exponential cost $L(t)$ defined in Eqn.~\eqref{eq:expcost}. In analogy with the classical case, we try to execute AC by dividing each segment according to the escort distribution $p^{(q)}$, where $p^{(q)}_i = \frac{p_i^q}{\sum_{j=1}^Np_j^q}$, in order to reach the optimal lengths defined in Eqn.~\eqref{eq:escortlen}. We will call this procedure AC$_q$. Moreover, we will call $\mathcal{S}_j^{(q)}$ the length of the segment generated by AC$_q$ at iteration $j$. The logarithm of the length of the final segment $\mathcal{S}^{(q)}_M$ for a string $\vec{s}$ is:
\begin{align}
\begin{split}
    \log_D \mathcal{S}^{(q)}_M(\vec{s}) & = \log_D \prod_{j=1}^M \frac{p_{i_j}^q}{\sum_{i=1}^N p^q_i} \\ & = \sum_{j=1}^M \bigg(\log_D p^q_{i_j} - \log_D \sum_{i=1}^N p_i^q  \bigg) \\
    &= q\sum_{i=1}^N n_i(\vec{s})\log_D p_i - M\log_D \sum_{j=1}^N p^q_j
\end{split}
\end{align}
where $n_i(\vec{s})$ counts how many times the symbol $x_i$ appears in the string $\vec{s}$. From this result, it is possible to evaluate the number of bits emitted to encode a particular string in a binary alphabet ($D=2$): \begin{align}
    \ell_2^{(q)}(\vec{s}) = \bigg\lceil \log_2 \frac{2}{\mathcal{S}^{(q)}_M(\vec{s})}\bigg\rceil < 2 - \log_2 \mathcal{S}^{(q)}_M(\vec{s}).
    \label{eq:lqsq}
    \end{align}
Let's define now the exponential cost $L_M(t)$ of a string of length $M$ composed by independent symbols: \begin{align}
\label{eq:expcostM}
    L_M(t) = \frac{1}{t}\log_2\sum_{\vec{s}}P(\vec{s})\: 2^{t\ell_2(\vec{s})},
\end{align}
where $P(\vec{s})=\prod_{i=1}^N p_i^{n_i(\vec{s})}=\mathcal{S}_M(\vec{s})$. Given Eqn.~\ref{eq:expcost}, it is $L_M(t) = M \cdot L(t)$. Substituting Eqn.~\eqref{eq:lqsq} in the definition of $L_M(t)$, and considering the optimal parameter value $q=1/(1+t)$, we get that:
\begin{align}
    \begin{split}
        L_M(t) &= \frac{1}{t}\log_2\sum_{\vec{s}}P(\vec{s})\: 2^{t\ell_2^{((t+1)^{-1})}(\vec{s})}\\
        &< \frac{1}{t}\log_2\sum_{\vec{s}}P(\vec{s})\: 2^{t(2 - \log_2 \mathcal{S}^{((t+1)^{-1})}_M(\vec{s}))}\\
        &= \frac{1}{t}\log_2 \Bigg(2^{2t}\: 2^{tM\log_2 \sum_j p_{i_j}^{(t+1)^{-1}}}\\
        &\;\;\;\;\;\;\;\;\;\cdot \sum_{\vec{s}}\bigg(P(\vec{s})\prod_{i=1}^N \big(p_i^{n_i(\vec{s})}\big)^{-\frac{t}{t+1}}\bigg)\Bigg)\\
        &= 2+\frac{Mt}{t+1}H_{\frac{1}{t+1}}[p]+\frac{1}{t}\log_2 \sum_{\vec{s}} P(\vec{s})^{1-\frac{t}{t+1}}\\
        &=2+\frac{Mt}{t+1}H_{\frac{1}{t+1}}[p]+\frac{M}{t+1}H_{\frac{1}{t+1}}[p] \\
        &= 2+MH_{\frac{1}{t+1}}[p].
    \end{split}       
\end{align}
Which reads:
    \begin{align}
    \label{eq:lt}
    L_M(t)< 2+MH_{\frac{1}{t+1}},
\end{align}
where $H_q[p] = \frac{1}{1-q} \log_2\sum_{i=1}^N p_i^q$ is the Rényi entropy of the source for a single symbol. Notice that for independent symbols the R\'enyi entropy is additive, i.e. for i.i.d. symbols $H_q[P]=M \cdot H_q[p]$ holds.
So, the compressor AC$_q$ leads to an average cost per symbol which is close to $H_{\frac{1}{1+t}}[p]$ as $M$ increases: 
\begin{align}
    L(t)=\frac{L_M(t)}{M}<\frac{2}{M}+H_{\frac{1}{1+t}}[p].
\end{align}
It is possible to visualize this result by considering the cost $L_M(t, q)$, in which the parameters $t$ and $q$ are now decoupled: $t$ is the exponent of the cost function, while $q$ is used in the AC$_q$ procedure. In particular, it reads: \begin{align}
\label{eq:expcostMqvar}
L_M(t,q)= \frac{1}{t}\log_2\sum_{\vec{s}}P(\vec{s})\: 2^{t\: \ell^{(q)}_2(\vec{s})}.
\end{align}
Here, $\ell^{(q)}_2$ represents the number of bits emitted by applying the AC$_q$ procedure with the escort distribution of order $q$ (notice that, if $q=1$, then the AC$_{q=1}$ reduces to the classic compressor AC). 

\begin{figure}[ht]
    \centering
    \includegraphics[scale=0.5]{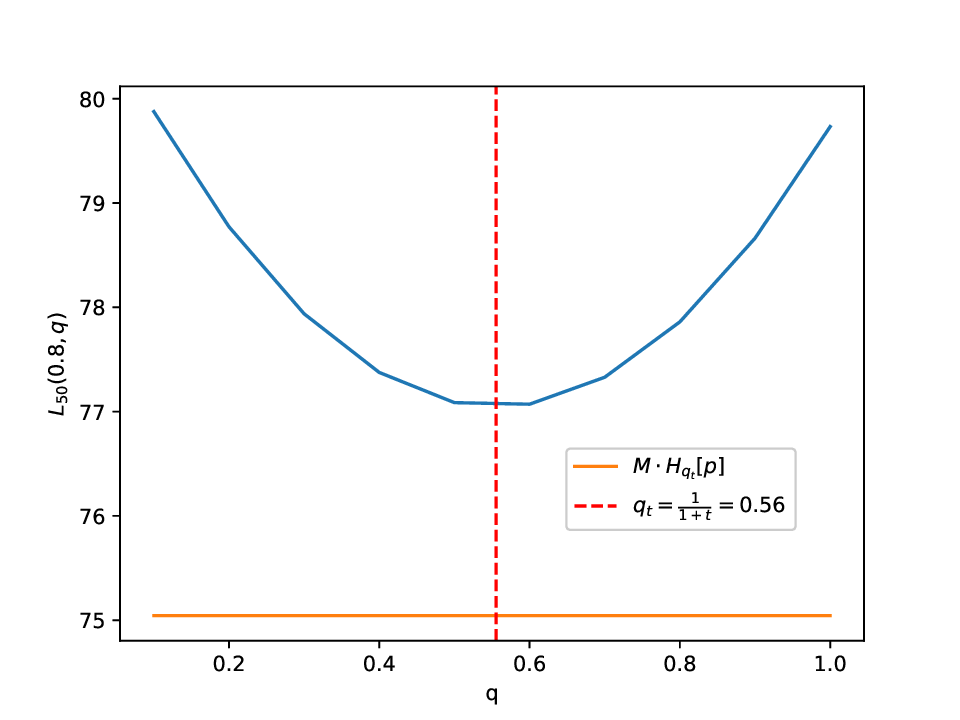}
    \caption{Exponential average codeword length of strings of length $M=50$, composed by i.i.d. symbols sampled according to $p_i\propto i^{-1}$, $i\in[1,3]$. Here $t=0.8$. Its minimum is reached in the proximity of the red vertical dashed line, corresponding to the optimal $q$, i.e. $q_t$. For that value of $q$, the distance with respect to the R\'enyi entropy of the string (flat orange line) is approximately $2$.}
    \label{fig:plot}
\end{figure}

Figure~\ref{fig:plot} shows  $L_M(t, q)$ for different values of $q$. The minimum is reached exactly in the value of $q$ prdicted by Campbell, that we now call $q_t=\frac{1}{1+t}$. Moreover, the distance between the minimum of $L_M(t, q)$, i.e. $L_M(t,q_t)$, and the orange line (corresponding to $M\cdot H_{\frac{1}{1+t}}[p]$) is very close to $2$, confirming the result of Eqn. (\ref{eq:lt}).

\subsection{A note on the semi-static approach}
In this section, we discuss how the probability distribution $p$ can be measured for different encoding schemes, focusing on the AC$_q$. An encoding scheme can be \textit{static} or \textit{semi-static}, depending on how the probability distribution of the source is computed or updated. In the first case, the probability distribution approximating the source's one is fixed and never changed while strings are generated. In the second case, instead, the probability distribution is evaluated each time a string needs to be encoded, and it is set equal to the frequency of symbols appearing in that string. In~\cite{moffat2020large}, the AMS coding is focused on the second case.

In this section, we want to elucidate how the use of a semi-static approach, instead of a static one, affects the exponential cost of Eqn.~\eqref{eq:expcostMqvar}. 

Let's assume that is possible to reach codewords' lengths as expressed in Eqn.~\eqref{eq:escortlen}, for any symbol and any value of $q$. Then, it is possible to write: \begin{align}
\begin{split}
    \ell_D^{(q)}(\vec{s})&=\sum_{j=1}^M\bigg(-\log_D \frac{p_{i_j}^q}{\sum_{i=1}^N p_i^q}\bigg) \\
    &= -\sum_{i=1}^N n_i(\vec{s})\log_D\frac{p_i^q}{\sum_{i=1}^N p_i^q} = M\cdot H_1[f(\vec{s})||p^{(q)}],
\end{split}
\end{align}
where $H_1[f\,||\,p]=-\sum_i f_i\log_D p_i$ is the \textit{cross-entropy} between distributions $f$ and $p$, and $f(\vec{s})=\big(\frac{n_1(\vec{s})}{M},\dots,\frac{n_N(\vec{s})}{M}\big)$ is the empirical frequency of each symbol in the string $\vec{s}$. We also remind that $p^{(q)}$ is the escort distribution of order $q$ of the distribution $p$. So, Eqn.~\eqref{eq:expcostMqvar} can be rewritten as: \begin{align} L_M(t,q)=\frac{1}{t}\log_D \bigg(\sum_{\vec{s}}P(\vec{s})\: D^{t\,M\,H_1[f(\vec{s})\,||\,p^{(q)}]}\bigg).
\end{align}
While Campbell~\cite{campbell1965coding} showed that the best strategy (i.e., the best $q$) to minimize the exponential cost consists of taking $q=q_t=1/(1+t)$, in the semi-static approach, the exponential cost of each string is minimized individually by taking $q=1$. The reason is that the cost of encoding a single string is $D^{tMH_1[f(\vec{s})||p^{(q)}]}$. Since it is assumed that the probability of the source is equal to the empirical frequency appearing in the string to encode, i.e. $p=f(\vec{s})$, setting $q=1$ provides the lowest cost (if $t>0$) since $$H_1[f^{(1)}(\vec{s})||f^{(1)}(\vec{s})]<H_1[f^{(1)}(\vec{s})||f^{(q)}(\vec{s})], \;\; \forall q\geq 0.$$ 

In other words, if one assumes that $p=f(\vec{s})$ then all the observed strings are encoded as if they were members of the \textit{typical} set of strings, thus they are better encoded by considering $q=1$, i.e. the Shannon-like approach \cite{cover1999elements}. Notice that if more than one string $\vec{s_i}$ is to be encoded, by assuming $p=f(\vec{s_i})$ at each string, one has to take into account that the probability distribution of the source is non stationary since, in general, $f(\vec{s}_i) \neq f(\vec{s}_j)$. This violates Campbell's hypothesis, thus making our compression approach non applicable in this case. 

On the other hand, the situation is much different if one considers the static approach. In this case, the strings $\vec{s}_i$ are considered to be generated by a stationary source according to a distribution $p$. So it becomes possible to observe strings whose corresponding $f(\vec{s})$ is outside the typical set of $p$, meaning that they are very expensive to encode and thus making the use of our approach very advantageous in the case of an exponential cost.

\section{Application to Wikipedia}
\label{sec:wikipedia}
Having delineated the theoretical side of our AC$_q$ in the preceding sections, we now transition to a more empirical scenario. This section is dedicated to the application of our outlined procedure to real-world data. 

\begin{figure}[ht]
    \centering
    \includegraphics[scale=0.5]{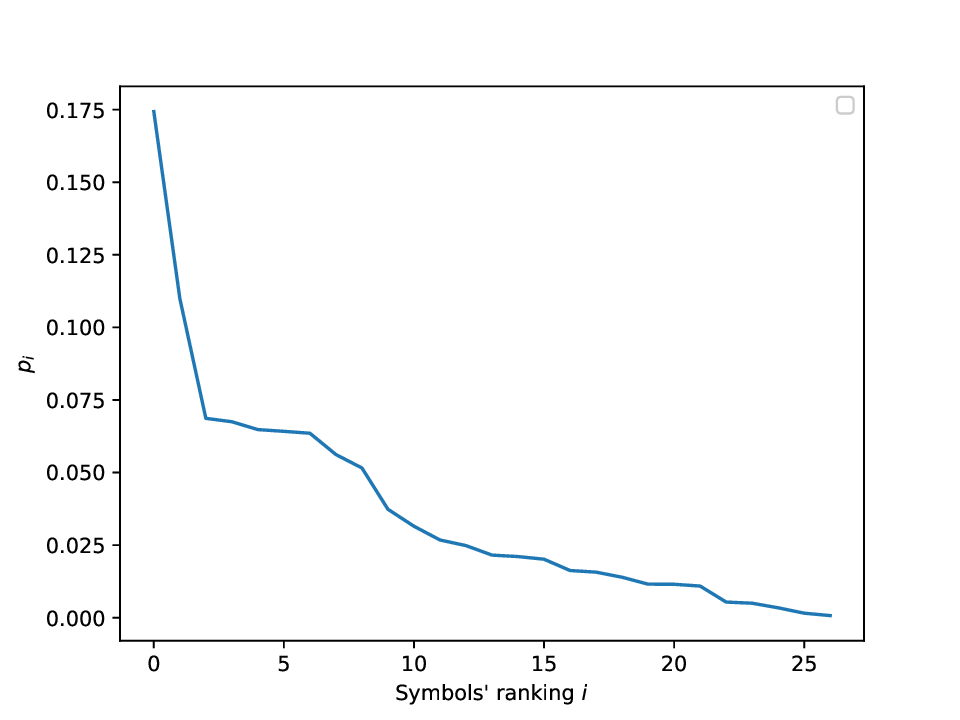}
    \caption{Probability distribution of the individual symbols (i.e., characters) in the Wikipedia dataset. Symbols have been ordered by decreasing frequency and assigned a rank. The probability distribution has been estimated in a frequentist approach as $p_i=n_i/W$, with $n_i$ being the number of times that symbol $x_i$ appears in Wikipedia.}
    \label{fig:wiki_prob_distr}
\end{figure}

In particular, we applied AC$_q$ to Wikipedia data.\footnote{The dataset FIL9 can be downloaded from \url{https://fasttext.cc/docs/en/unsupervised-tutorial.html}.} The dataset used for our analysis contains  $W\approx 7\cdot 10^8$ symbols from an alphabet $\Sigma$ of size $|\Sigma|=N=27$. In order to perform coding in a static approach as we mentioned earlier, we computed from the whole dataset the empirical frequency of the 27 distinct symbols, shown in Fig.\ref{fig:wiki_prob_distr},  and then used it to set the probability distribution $p=\{p_1,\dots,p_{27}\}$.

Since the theoretical results presented so far are valid for i.i.d. symbols, we first discuss and apply our procedure in the case of i.i.d. symbols. After that, we will move to the real Wikipedia dataset. 

To begin with, we generated $\eta=3.500.000$ strings of length $M=20$ composed by i.i.d. symbols sampled according to $p$. We then applied the AC$_q$ for different and discretized values of $q$. For each string we evaluated the length of the corresponding codeword generated by AC$_q$ algorithm, without actually generating it, as $\ell_2^{(q)}(\vec{s}) = \log_2 \lceil \frac{2}{\mathcal{S}^{(q)}_M(\vec{s})}\rceil$. Such lengths have been stored in a matrix $\mathcal{L}$, whose entry $\mathcal{L}_{ij}$ is the length of the codeword of the $j$-th string, generated with AC$_{q_i}$, where $q_i$ is the $i$-th value of $q$ that we encode with. Algorithm~\ref{alg:wiki} summarizes this procedure.
\begin{algorithm}[ht]
\caption{Wikipedia data analysis}\label{alg:wiki}
\begin{algorithmic}[1]
\REQUIRE data, $\Sigma=(x_1,\dots,x_N)$, $p = (p_1,\dots,p_N)$\\
\ENSURE $\text{len}(data)=M\cdot \eta$
\ENSURE $q_{end}>0$
\STATE $M \gets 20$
\STATE $\eta\gets\frac{\text{len}(data)}{M}$
\STATE $q\gets 0$
\STATE $q_{idx}\gets 0$
\WHILE{$q \leq q_{end}$}
\STATE  $n\gets0$
\STATE $p_i^{(q)}\gets \frac{p_i^q}{\sum_j p_j^q}$  \space $ \forall x_i \in \Sigma$
\WHILE{$n < \eta$}
\STATE $\text{string} \gets \text{data}[M\cdot n:M\cdot(n+1)-1]$
\STATE $\mathcal{S} \gets 1$
\FOR{c \text{in string}}
\STATE $i^*\gets i | c=x_i$ 
\STATE $\mathcal{S} \gets \mathcal{S}\cdot p_{i^*}^{(q)}$
\ENDFOR
\STATE $\mathcal{L}[q_{idx}, n] \gets \log_2 \lceil \frac{2}{\mathcal{S}}\rceil $
\STATE $n \gets n+1$
\ENDWHILE
\STATE $q \gets q+0.1$
\STATE $q_{idx} \gets q_{idx}+1$
\ENDWHILE
\end{algorithmic}  
\end{algorithm}
By using the matrix $\mathcal{L}$ generated by Algorithm~\ref{alg:wiki}, it is possible to evaluate the empirical exponential average length, for different values of $t$, as: \begin{align}
   L_M^{emp}(t,q_i) = \frac{1}{t}\log_2\Bigg(\frac{1}{\eta}\sum_{j=1}^{\eta} 2^{t\mathcal{L}_{ij}} \Bigg). 
\end{align}

\begin{figure}[ht!]
    \centering    \includegraphics[scale=0.5]{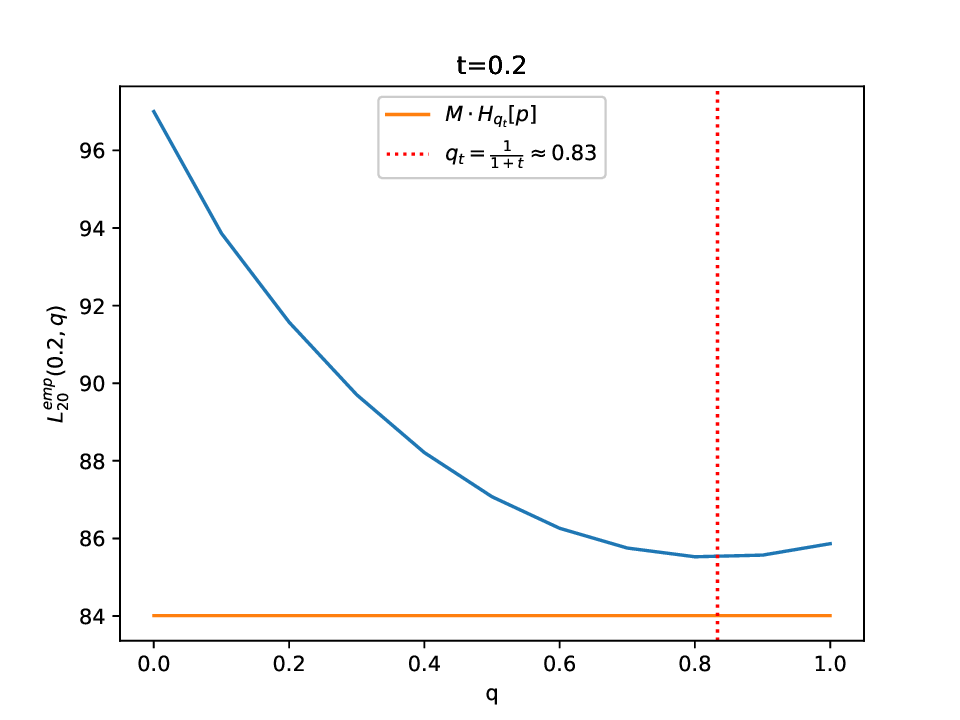}
    \includegraphics[scale=0.5]{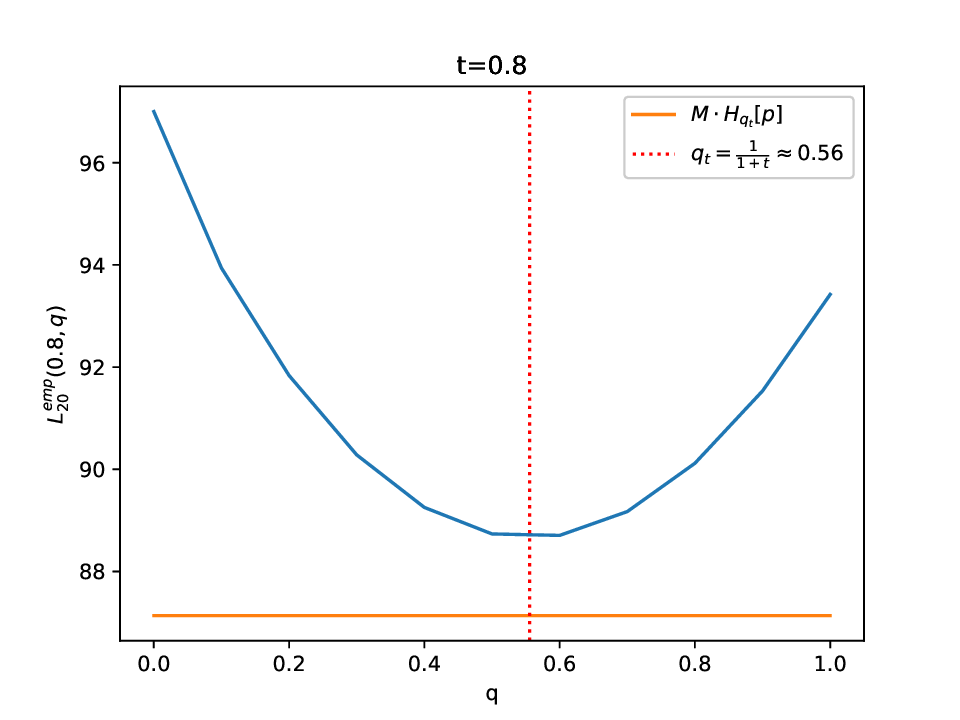}
    \includegraphics[scale=0.5]{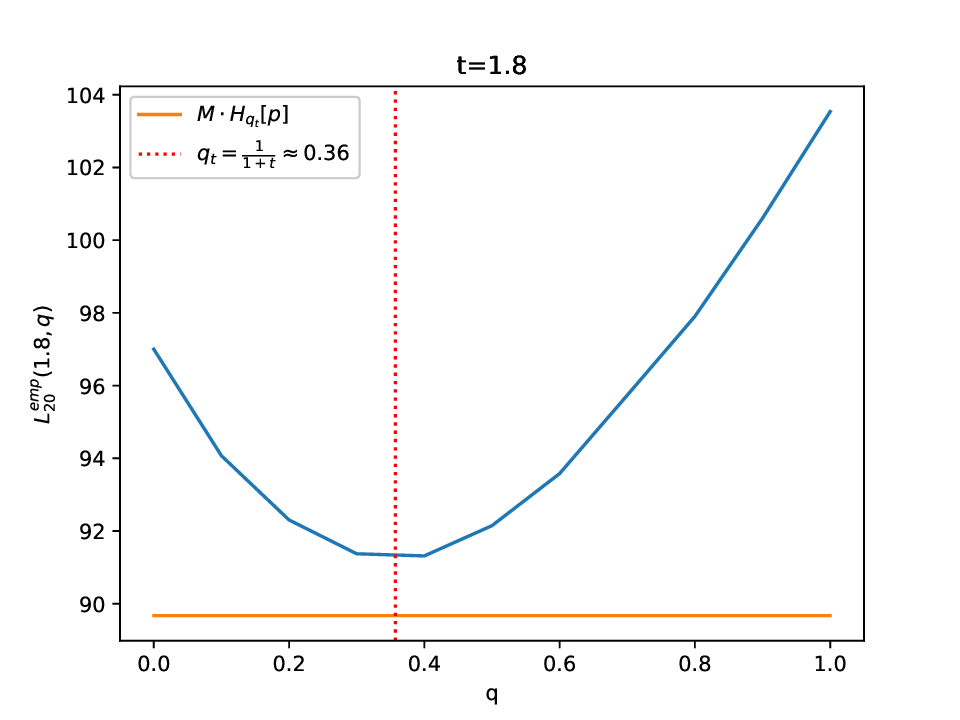}
    \caption{This Figure shows, for the synthetic i.i.d. generated symbols, the empirical exponential average $L_M^{emp}(t,q)$ for $M=20$ (blue line), the R\'enyi entropy $M\cdot H_{q_t}[p]$ (horizontal orange line) with $q_t=1/(1+t)$ (dotted vertical red line). The three panels show different values of $t\in\{0.2,0.8,1.8\}$. We can see that, in each case, the minimum of $L_M^{emp}(t,q)$ is reached at $q=q_t$, and that $L_M^{emp}(t,q_t)- M\cdot H_{q_t}[p] \approx 2$.} 
    \label{fig:emp_exp_cost_iid}   
\end{figure}
Figure~\ref{fig:emp_exp_cost_iid} shows the empirical $L_M^{emp}$ as a function of $q$ for three different values of $t$, with the corresponding Rényi entropy $M\cdot H_{q_t}$ and the optimal $q_t=1/(1+t)$. It is clear that the minimum of $L_M^{emp}(t,q)$ is reached at $q=q_t$ and that it is very close to the R\'enyi entropy $M\cdot H_{q_t}[p]$.

\begin{figure}[ht!]
    \centering    \includegraphics[scale=0.5]{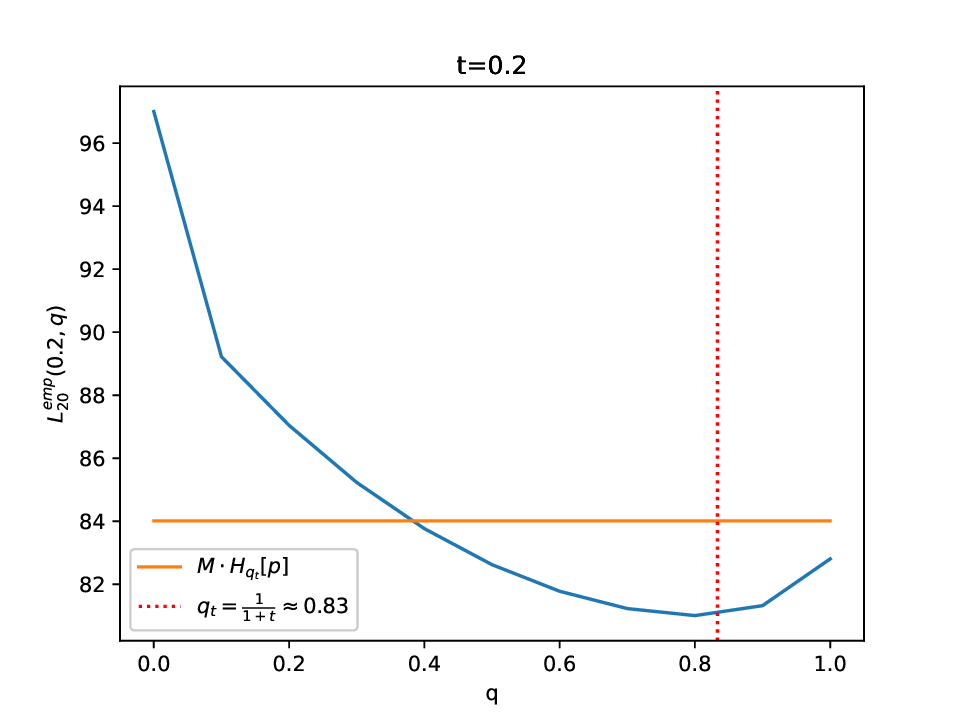}
    \includegraphics[scale=0.5]{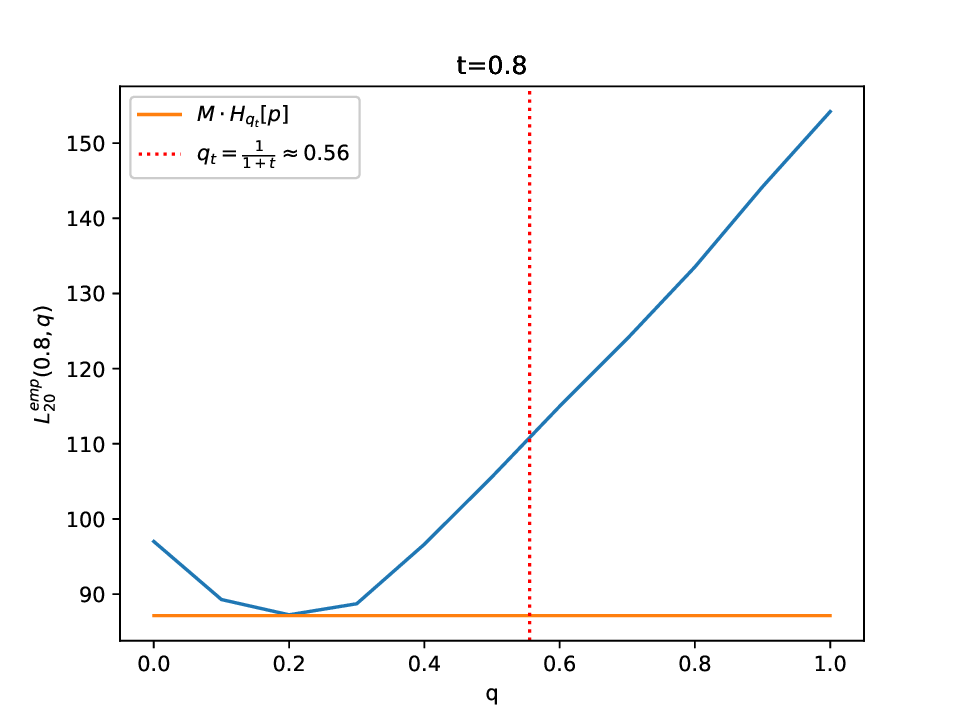}
    \includegraphics[scale=0.5]{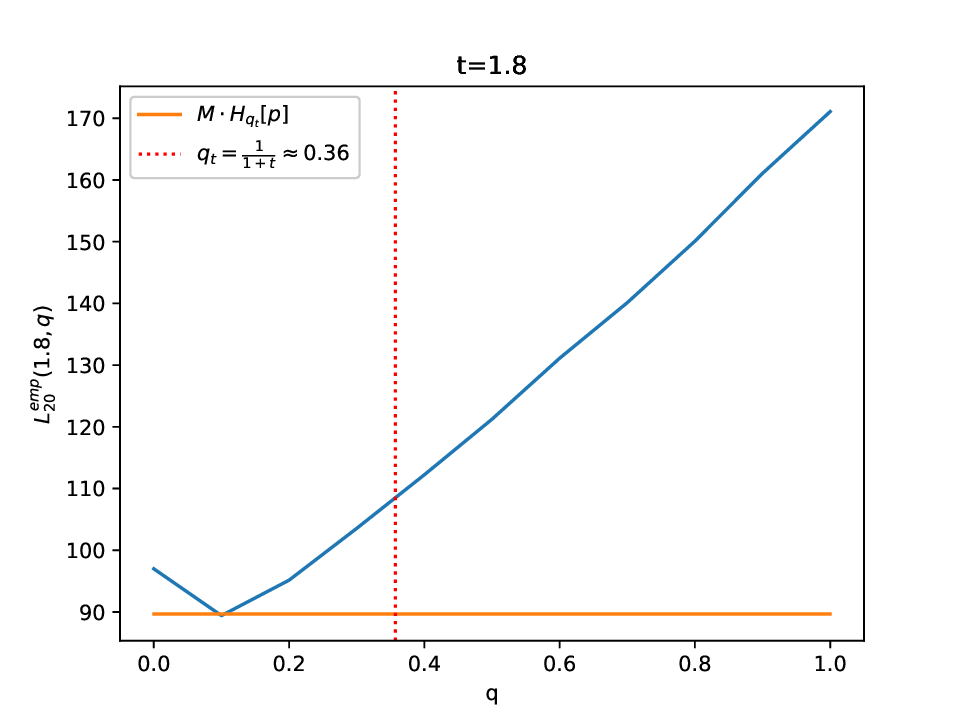}
    \caption{This Figure shows, for the real Wikipedia dataset,, the empirical exponential average $L_M^{emp}(t,q)$ for $M=20$ (blue line), the R\'enyi entropy $M\cdot H_{q_t}[p]$ (horizontal orange line) with $q_t=1/(1+t)$ (dotted vertical red line). The three panels show different values of $t\in\{0.2,0.8,1.8\}$. In all these cases, it is instead evident that the minimum of $L_M^{emp}(t,q)$ is reached for $q < q_t$. Additionally, $\min_q L_M^{emp}(q,t)$ could be lower (first panel) or almost exactly reach (second and third panels) the value $M\cdot H_{q_t}[p]$. We can also see that encoding according to AC$_{q_t}$ (i.e., see the intersection between the blue line and the dotted line) can lead to an exponential average length smaller than the R\'enyi entropy (first panel), or to an error which greater than $2$, i.e. $\min_q L_M^{emp}(t,q)-M\cdot H_{q_t}[p]>2$ (second and third panels).}
    \label{fig:emp_exp_cost}
\end{figure}

Let us now move to analyze the real Wikipedia dataset. We divided it in $\eta=35.653.488$ strings of length $M=20$. We applied again Algorithm~\ref{alg:wiki}, and proceeded in the same way as we did for the i.i.d. symbols scenario. Figure~\ref{fig:emp_exp_cost} reports our results. In particular, we can see that $\mathrm{argmin}_q (L_M^{emp}(t,q))<q_t$ and that, for $t=0.2$ (top panel), our AC$_q$ can perform better than what Campbell predicted (in fact, $\min_q L_M^{emp}(t,q)<M\cdot H_{q_t}[p]$). The emergence of such discrepancies is not surprising, since real English text is not composed by i.i.d. symbols, and thus the hypothesis on which our theoretical description lies is not satisfied. 

\begin{figure}[ht!]
    \centering
    \includegraphics[scale=0.5]{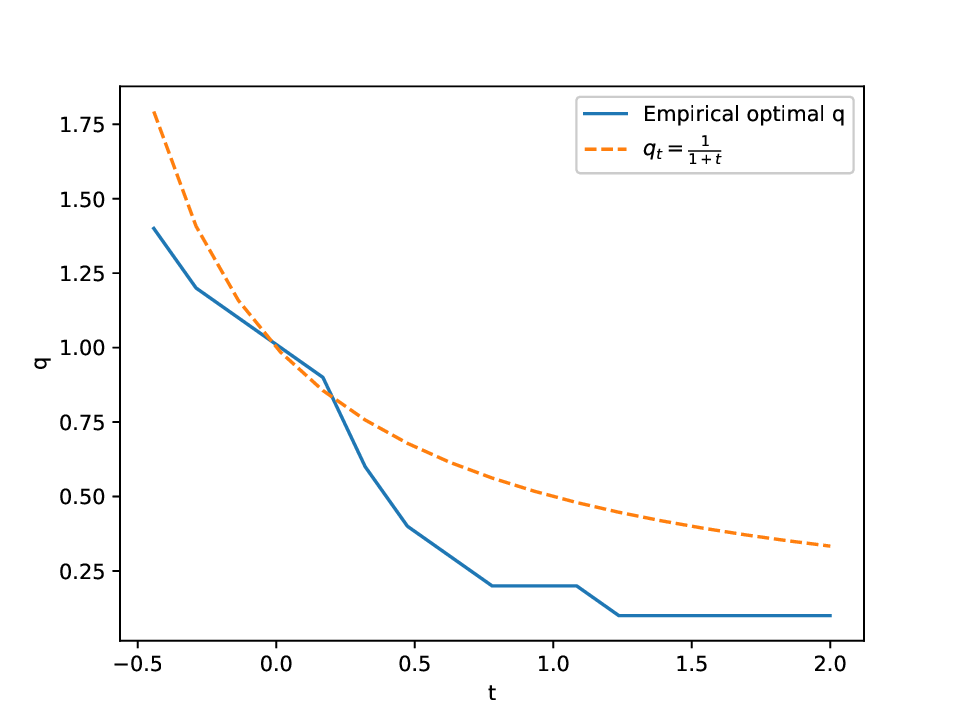}
    \caption{Empirical best $q$ for the Wikipedia dataset (blue line solid line) and $q_t$ (orange dashed line) for different values of $t$. For almost every value of $t$, the empirical optimal $q$ is smaller than $q_t$, meaning that in the Wikipedia dataset there is an abundance of `rare' strings.}
    \label{fig:qthvsqempwiki}
\end{figure}

But what does it mean, `physically', the fact that, in this case, the average empirical cost is minimized by considering a $q$ smaller than $q_t$? Since we are using escort distributions $p^{(q)}$ of order $q$ as encoding strategy in Eqn.~\eqref{eq:escortlen}, decreasing the value of $q$ is equivalent to increasing the probability of the rare strings. This translates into assigning them shorter codewords, more than it would be done by using $q=q_t$. In other words, when the real optimal $q$ is smaller than $q_t$, this means that `rare' strings are actually more abundant in the dataset than they would be if they were generated by a probability distribution calculated as the product of the probability of i.i.d. symbols. Figure~\ref{fig:qthvsqempwiki} shows, for different values of $t$, the real (empirical) optimal $q$ overlapped to the theoretical $q_t$. For most values of the exponent $t$, the empirical best $q$ is smaller than $q_t$. 

\begin{figure}[ht]
    \centering
    \includegraphics[scale=0.5]{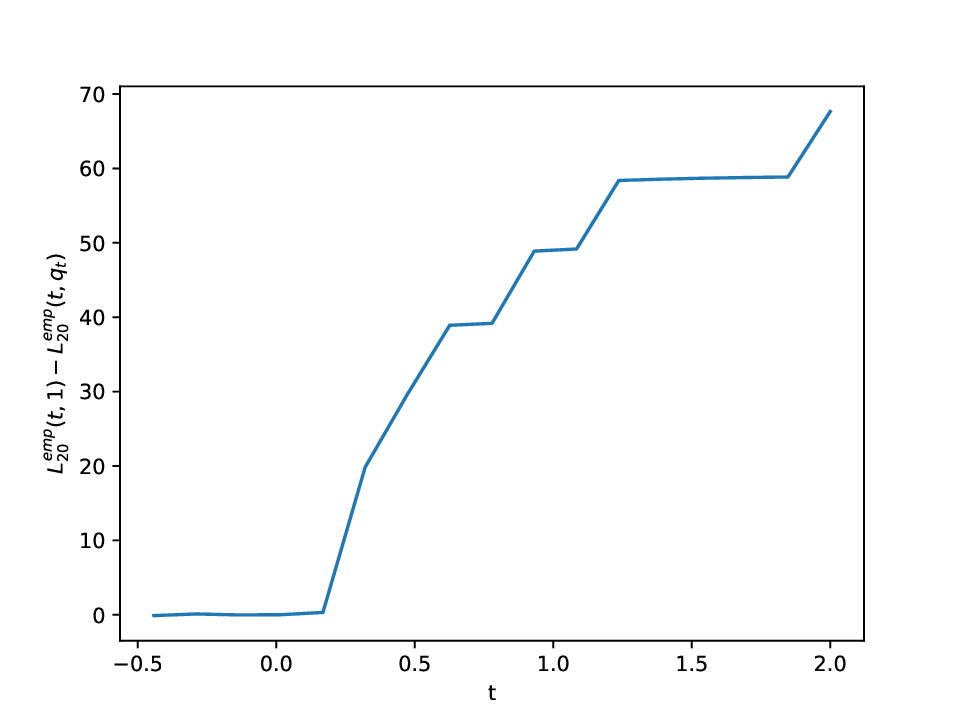}
    \caption{Difference $L_M^{emp}(t,q=1)-L_M^{emp}(t,q=q_t)$ for $M=20$ as a function of $t$, for the Wikipedia dataset. Since, for $t$ big enough, this quantity is positive (and increasing), if the average to consider is the exponential one, there is an advantage (which increases with $t$) in encoding according to $q_t$ instead of $q=1$.}
    \label{fig:cost_diff}
\end{figure}

Finally, we want to stress that even if English text does not satisfy the i.i.d. symbols hypothesis on which Campbell's theoretical description lies, the use of the AC$_q$ still outperforms the standard AC if the average length is exponential, although the empirical optimal $q$ is not the one predicted by Campbell. In fact, while the value of $q$ that is actually optimal in the case real English text can not be known a priori, by using the one which is optimal for i.i.d. symbols (i.e. $q_t$) it is possible to significantly reduce the exponential average length, or the cost, with respect to the standard case $q=1$. This is shown in Figure~\ref{fig:cost_diff}, where we can see that, even if the true optimal $q$ is different from $q_t$, by encoding according to $q_t=\frac{1}{1+t}$ there is a notable exponential average length drop with respect to the usual $q=1$ encoding strategy. Of course, if one would know the true optimal $q$ the advantage would be even greater.

\section{Discussion}
\label{sec:discussion}
In this section, we'll take a closer look at the core ideas and findings from our research. We'll first explore one of the reasons behind the use of the exponential cost in our study, thus explaining why it's important and how it fits into the bigger picture of data compression. After that, we will provide further analysis of real data, which confirms our previous findings. Moreover, we will also mention what are the errors that can come up when our guesses about the true probability distribution of the source are not accurate, shedding light on some of the challenges we faced and how they might be addressed in future studies.

\subsection{A justification to the exponential cost with Cramér's theorem}
\label{sub:cramer}
In this subsection, we provide a simple yet powerful idea about the usefulness of the exponential average and its minimization. Such idea relies on the linkage between the exponential average and the cumulant generating function of a distribution. As we anticipated in the introduction of this paper, such application could be useful in scenarios in which it is imperative to minimize the probability that codewords' lengths exceed a certain threshold.

Suppose that we are interested in encoding strings of fixed length $M$, and that we do not want the corresponding codewords length's exceed the threshold $M\cdot a$. So, using the usual notation, the aforementioned problem translates into finding an encoding strategy $x_i \rightarrow \ell_D(i)$ that assigns to each symbol $x_i$ of the alphabet $\Sigma$ a length $\ell_D(i)$ to its codeword minimizing: 
\begin{align}
        \text{Prob}\Bigg[\frac{1}{M}\sum_{j=1}^M \ell_D(i_j) \geq a\Bigg].
        \label{ex:pa}
\end{align}
Here, the sum runs over the whole string, and $\ell_D(i_j)$ is the length of the encoded symbol appearing at the $j$-th position of the string to be encoded. Moreover, since the threshold for the encoded strings is $M\cdot a$, we can see $a$ as the threshold \textit{per symbol}.
According to Cramér's theorem, it is possible to write the following Chernoff bound:
\begin{align}
\label{eq:cramer}
    \text{Prob}\Bigg[\frac{1}{M}\sum_{j=1}^M \ell_D(i_j) \geq a\Bigg] \leq e^{-M\big(ta-\mu(t)\big)} \;\; \forall t>0,
\end{align}
where $\mu(t)=\log \mathbb{E}_p[e^{t\ell_D(i)}]$ is the symbols' distribution's cumulant-generating function and $\log$ is the natural logarithm (i.e. with base $e$). Eqn.~\eqref{eq:cramer} gives us an important degree of control on the probability of exceeding the threshold, since, as we will show, it is possible to control its upper bound. Of course, we are interested in situations in which the exponent $-M\big(ta-\mu(t)\big)$ is negative, otherwise, we would get an upper bound of a probability distribution greater than $1$, thus totally uninformative. It is possible to rewrite the exponent of the upper bound as:
\begin{align}
\begin{split}    
\label{eq:exponent}
    -M(ta-\mu(t)) &= -M\Bigg(ta-\log\big(\sum_i p_i e^{t\ell_D(i)}\big)\Bigg) \\
    &=-M\Bigg(ta-\frac{t\log_D(\sum_i p_i D^{t\ell_D(i)\log_D e})}{t\log_D e}\Bigg)\\
    &= -M\cdot t(a-L(t\log_D e)).
\end{split}
\end{align}
Since $M$, $t$ and $a$ are positive by definition, we are interested in finding the strategy minimizing $L(t\log_D e)$ $\forall t>0$. We know that, for a given value of $t' =t\log_D e$, the minimum of $L(t')$ is $H_{q_{t'}}[p]$, with $q_{t'}=\frac{1}{1+t'}$. Moreover, we know that such minimum is reached with the strategy $\ell_D^{(q)}(i) = -\log_D p^{(q)}_i$ (see Eqn.~\eqref{eq:escortlen}). So, by writing Eqn.~\eqref{eq:exponent} as a function of $q_{t'}$ (and, for simplicity, by dropping the subscript `$t'$'), we get that:
\begin{align}
\label{eq:exp_renyi}
    -M(ta-\mu(t)) =-M \frac{1-q}{q\log_D e}(a-H_{q}[p]).
\end{align}
So, it is possible to write Eqn.~\eqref{eq:cramer} as:
\begin{align}
\label{eq:cramerren}
    \text{Prob}\Bigg[\frac{1}{M}\sum_{j=1}^M \ell_D(i_j) \geq a\Bigg] \leq e^{-M \frac{1-q}{q\log_D e}(a-H_{q}[p])} \;\; \forall q\in (0,1].
\end{align}
Having pointed out that the best strategy consists in setting the codewords lengths according to Eqn.~\eqref{eq:escortlen}, with $q=q_{t'}=1/(1+t')$, we have to determine which is the correct $t>0$ (and, in turn, $q_t$) to consider. We expect that the choice depends on the threshold $a$. In order to choose the best parameter $q$, we will minimize the right-hand side of Eqn.~\eqref{eq:exp_renyi}. Since we are assuming it to be negative, this guarantees that the upper bound in Eqn.~\eqref{eq:cramerren} is minimized. 

Before going into the analytical details of such minimization, we will consider two simple examples which will provide an intuition on how the encoding strategy is related to the threshold $a$. Recall that  $H_q[p]$ is a decreasing function of $q$, i.e., $H_0[p] \geq \dots \geq H_1[p]$.
\paragraph{Case $a>H_0[p]=\log_D|\Sigma|$.}
In the first case we consider, we assume that the threshold $a$ is bigger than the R\'enyi entropy of order $0$. Since $H_0[p]=\log_D |\Sigma|$, we are assuming that the threshold exceeds the Shannon entropy of a distribution that shares the same support as the original $p$, but with entries replaced by $1/|\Sigma|$, i.e. a uniform distribution. In this scenario, the term $(a-H_q[p])$ in Eqn.~\eqref{eq:exp_renyi} is positive and finite $\forall q\in(0,1]$. The r.h.s. of Eqn.~\eqref{eq:exp_renyi} is then maximized by letting $q \to 0$ (i.e. $t\to
+\infty$). By writing $a= H_0[p] + \epsilon$, with $\epsilon>0$, Eqn.~\eqref{eq:cramerren} reads:
\begin{align}
    P\Bigg[\frac{1}{M}\sum_{j=1}^M \ell_D(i_j) \geq H_0[p]+\epsilon\Bigg] \leq \lim_{q\to 0} e^{-M \frac{1-q}{q\log_D e}\epsilon}=0.
\end{align}
So, the probability of emitting a codeword longer than the threshold vanishes. This result is trivial: by setting $q\to 0$, the encoding strategy is equivalent to the Shannon encoding for symbols generated with a uniform probability distribution. In fact, $\ell_D^{(0)}(i) = -\log p^{(0)}_i=\log|\Sigma|$, and this holds for any probability distribution $p$.  In other words, if it is imperative that the average codeword length does not exceed $H_0[p]$, just encode the sequence as if the symbols are uniformly distributed, irrespective of their actual probability distribution.

\paragraph{Case $a<H_1[p]$.}
In this second case, we are going to consider a threshold smaller than the Shannon entropy of the underlying probability distribution $p$. So, it follows that $(a-H_q[p])$ is negative $\forall q$ because $H_0[p] \geq \dots \geq H_1[p] > a$. Then, by setting $a=H_1[p]-\epsilon$, with $\epsilon>0$, Eqn.~\eqref{eq:cramerren} reads:
\begin{align}
    P\Bigg[\frac{1}{M}\sum_{j=1}^M \ell_D(i_j) \geq H_1[p]-\epsilon\Bigg] \leq e^{-M\frac{1-q}{q\log_d e}\big(H_1[p]-\epsilon-H_q[p]\big)}.
\end{align}
The exponent $-M\frac{1-q}{q\log_d e}\big(H_1[p]-\epsilon-H_q[p]\big)>0$ is positive $\forall M \in \mathbb{N}$, and so it does not satisfy our hypothesis of a negative exponent. As previously mentioned, this means that the above right-hand side term is greater than $1$. For this reason, it gives no information on the probability of exceeding the threshold. We can however see that since $H_1[p]$ is the shortest achievable codewords' (linear) average length, the latter can be smaller than $H_1[p]$ only due to fluctuations in the observed symbols frequency, which are suppressed in the large $M$ limit. The best strategy is then letting $q=1$, but still, the threshold will be exceeded almost always if $M$ is not unrealistically small.

\paragraph{Case $H_1[p]\leq a \leq H_0[p]$.} 
Now that we have shown the two extreme cases $a>H_0[p]$ and $a<H_1[p]$, let's focus our attention on the most interesting case: i.e. $H_0[p]\leq a\leq H_1[p]$. As previously mentioned, we are interested in finding the value $q=q^*$ for which $-M\frac{1-q}{q\log_D e}(a-H_q[p])$ is minimized. Taking the derivative, one gets:
\begin{align}
\begin{split}  
    \label{eq:der}
    \frac{d}{dq}\Bigg(&-M\frac{1-q}{q\log_D e}(a-H_q[p])\Bigg)= \\
    &=-\frac{M}{\log_D e}\Bigg(\frac{1}{q(1-q)}D_{KL}(p^{(q)}||p)-\frac{1}{q^2}(a-H_q[p])\Bigg),
    \end{split}
\end{align}
where $D_{KL}(p^{(q)}||p)=\sum_{i}p^{(q)}_i\log_D \frac{p^{(q)}_i}{p_i}$ is the Kullback-Leibler divergence between the escort of $p$ and $p$ itself. The minimum is then found by setting the derivative to zero, leading to the condition:
\begin{align}
\begin{split} 
    \label{eq:der1}
    a-H_{q^*}[p] &= \frac{q^*}{1-q^*}D_{KL}(p^{(q^*)}||p)\\
    &= \frac{q^*}{1-q^*}\sum_{i=1}^{|\Sigma|} p_i^{(q^*)}\log_D \frac{p_i^{(q^*)}}{p_i}\\
    &=\frac{q^*}{1-q^*}\Bigg[\sum_{i=1}^{|\Sigma|}\Bigg(\frac{p_i^{q^*}}{\sum_{j=1}^{|\Sigma|}p_j^{q^*}}\log_D p_i^{q^*-1}\Bigg)\\
    &\;\;\;\;\;\;\;\;\;\;-\sum_{i=1}^{|\Sigma|}\Bigg(\frac{p_i^{q^*}}{\sum_{j=1}^{|\Sigma|}p_j^{q^*}}\log_D \sum_{j=1}^{|\Sigma|}p_j^{q^*}\Bigg)\Bigg]\\
    &= q^*(H_1[p^{(q^*)}||p]-H_{q^*}[p]).
    \end{split}
\end{align}
Moreover, it is useful to write the Shannon entropy of the escort $p^{(q)}$:
\begin{align}
    \begin{split}\label{eq:shannonofescort}
    H_1[p^{(q)}]&=-\sum_{i=1}^{|\Sigma|}\frac{p_i^q}{\sum_{j=1}^{|\Sigma|}p_j^q}\log_D \frac{p_i^q}{\sum_{j=1}^{|\Sigma|}p_j^q}\\
    &=qH_1[p^{(q)}||p]+(1-q)H_q[p]
    \end{split}
\end{align}
By plugging Eqn.~\eqref{eq:shannonofescort} into Eqn.~\eqref{eq:der1}, one gets that the value $q^*$ which sets the derivative to $0$ (i.e. minimizes the upper bound in the r.h.s. of Eqn.~\eqref{eq:cramerren}) satisfies:
\begin{align}
    \label{eq:der0}
    H_1[p^{(q^*)}]=a.
\end{align}
This equation relates the threshold $a$ to the encoding strategy driven by  $p^{q^*}$. In particular, such relation unveils that, if $a\in[H_1[p], H_0[p]]$, the optimal encoding strategy `pretends' that the symbols are generated according to their distribution's escort instead of the original $p$. Then, since by encoding with AC$_{q^*}$, we are actually (almost) reaching the shortest linear average length if symbols were generated according to $p^{(q^*)}$, it is reasonable that the best $q=q^*$ is the one for which the threshold is such shortest linear average, i.e. $H_1[p^{(q^*)}]$.

\smallskip Summarizing our contributions in this section, we note that we have justified the use of the exponential average by the necessity of not exceeding a certain threshold in the length of the encoded string. In particular, given the value of the threshold $a$ as an input, the procedure has three steps: 
\begin{enumerate}
    \item Estimate the probability distribution $p$ of the input symbols.
    \item Find $q^*$ by solving Eq.~\eqref{eq:der0}.
    \item Encode the input data with AC$_{q^*}$.
\end{enumerate}
Such procedure guarantees that, if $a>H_1[p]$, it is possible to reduce the number of codewords exceeding the threshold with the use of the described AC$_q$ algorithm, which reaches the R\'enyi entropy bound with an error of at most $2$ bits.

In the following paragraph, we will show a couple of examples over real and simulated data on how to infer the proper $q^*$, and how much this choice impacts the fraction of strings exceeding the threshold.

\subsection{Example}
Throughout this section, we will apply our procedure to both the usual Wikipedia dataset and simulated strings composed by i.i.d. symbols. We will generate the latter according to the probability $p=(p_1, \dots, p_{27})$ extracted from the Wikipedia dataset (see Figure~\ref{fig:wiki_prob_distr} for a visual reference). 
In order to understand which is the range of interest for the threshold $a$, we have evaluated that $H_0[p]\approx 4.75$ and $H_1[p] \approx 4.12$. For this reason, we will consider a threshold $a\in [4.12, 4.75]$.
\begin{figure}[ht]
    \centering
    \includegraphics[scale=0.5]{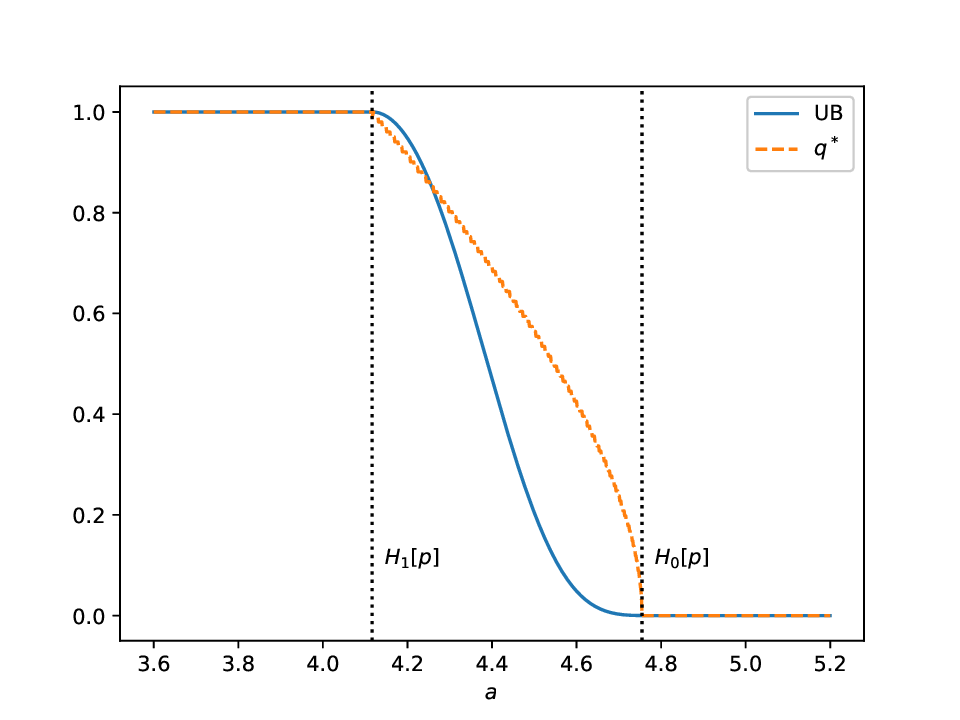}
    \caption{Dashed orange line: $q^*$ as solution of Eqn.~\eqref{eq:der0}. Solid blue line: upper bound of the probability of exceeding the threshold. Black dotted vertical lines: $H_1[p]$ and $H_0[p]$. As the threshold $a$ increases, the probability of exceeding it decreases until it reaches $0$ for $a=H_0[p]$.}
    \label{fig:q_star}
\end{figure}
Figure~\ref{fig:q_star} shows both the value $q^*$ for different values of $a$, evaluated as the solution of Eqn.~\eqref{eq:der0}, and the corresponding upper bound ($\mathrm{UB}$) of the probability of exceeding the threshold, evaluated as: \begin{align}
    \mathrm{UB}= e^{\big(-M \frac{1-q^*}{q^*\log_D e}(a-H_{q^*}[p])\big)},
\end{align}
where we set $M=20$. For $a\leq H_1[p]$, the upper bound $\mathrm{UB}$ is equal to $1$, thus it gives no information on the probability of exceeding the threshold. Instead, when $a$ increases, $\mathrm{UB}$ gets smaller until, for $a\geq H_0[p]$, it reaches $0$ (and so does $q^*$),  meaning that if the threshold is bigger than $H_0[p]$, by encoding with escort distribution of order $0$ it becomes impossible to exceed the threshold. This agrees with our previous analysis. 

\begin{figure}[ht]
    \centering 
    \includegraphics[scale=0.5]{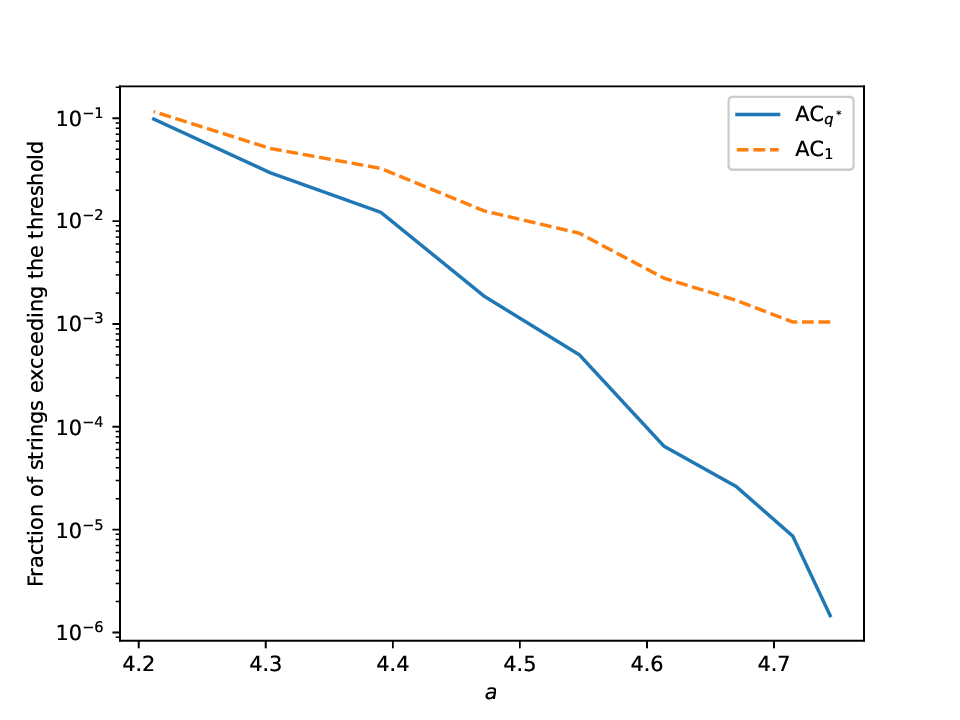}
    \includegraphics[scale=0.5]{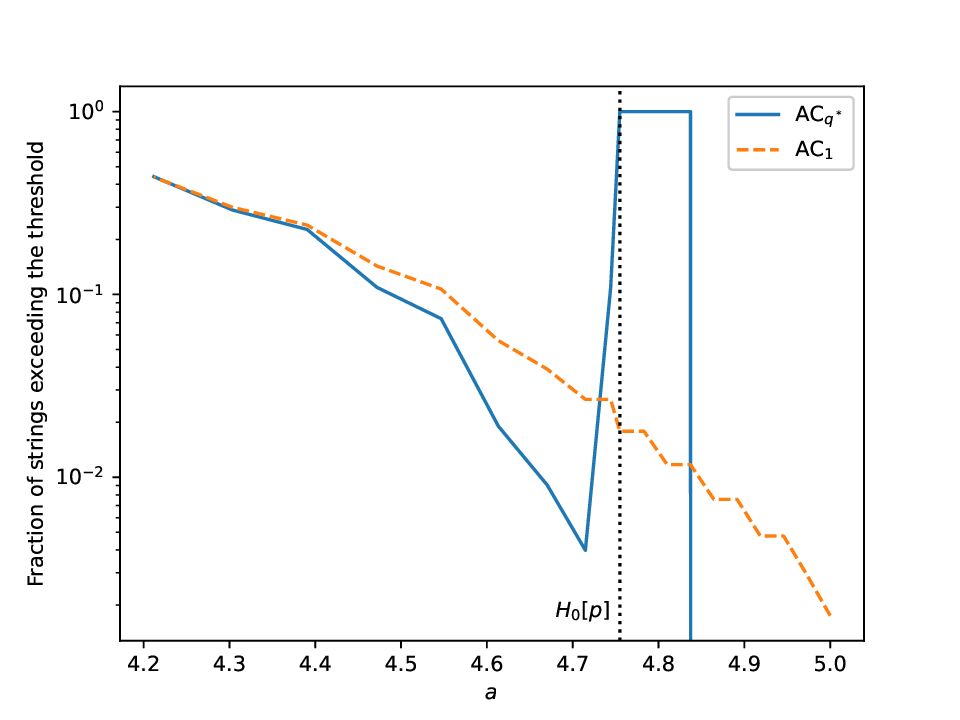}
    \caption{Fraction of strings exceeding the threshold for Wikipedia data (top panel) and for the simulated i.i.d. symbols (bottom panel). The orange dashed line is obtained with the classic arithmetic coder AC$_1$, while the solid blue line with AC$_{q^*}$. In the second panel, the spike before the plateau comes from the fact that an error $2/M=2/20=0.1$ occurs in AC$_q$ procedure. Indeed, the plateau's width is approximately $0.1$.}
    \label{fig:frac_exceeding}
\end{figure}

So, we expect that, by applying AC$_{q^*}$ to both Wikipedia and simulated data, the fraction of strings that exceed the threshold $M \cdot a$ is smaller than the one obtained by using the classic arithmetic coder, i.e. AC$_1$. Figure~\ref{fig:frac_exceeding} shows, as a function of $a$, the fraction of strings of length $M=20$ exceeding the threshold $M\cdot a$ when AC$_{q^*}$ and AC$_1$ are applied, over Wikipedia and simulated data (i.i.d. symbols). It can be noted that, by generalizing the encoding procedure, the number of codewords exceeding the threshold can be decreased significantly, especially for `large' $a$. Such a drop is more pronounced in the case of the Wikipedia data. The reason is that, since there is an abundance of `rare' strings in the real data (as we already discussed), the encoding strategy with escort distribution, which penalizes frequent symbols in favor of rare ones, is more efficient than it is for truly i.i.d. symbols. 

Moreover, we also want to explain the presence of the spike, followed by a short plateau, in the fraction of strings exceeding the threshold shown in the bottom panel of Figure~\ref{fig:frac_exceeding}, occurring for $a\gtrsim H_0[p]$. It is caused by the intrinsic 2-bits error of AC$_{q^*}$ procedure (see Eqn.~\eqref{eq:lt}). In fact, if $a=H_0[p]\approx 4.75$, then $M\cdot a \approx 95$. If we could exactly reach the desired symbols' length of Eqn.~\eqref{eq:escortlen} with $q^*=0$, we would never exceed the threshold. But AC$_{q^*}$ carries an intrinsic error: the encoded strings' lengths are all $97$ bits, in accordance with the predicted AC$_{q^*}$ error. The fraction of strings exceeding the threshold is then $1$ until $a$ becomes such that $M\cdot a =97$, i.e. $a=4.85$. After such value, the exceeding fraction drops to $0$. In other words, when the threshold is close to $H_0[p]$, even the very small error of the Arithmetic Coding procedure can lead to exceed it. Despite that, the ensemble of such cases is very small with respect to all the possibilities: for every $a\in (4.12, 4.75)$ the AC$_{q^*}$ procedure performs better than the usual AC$_1$, both for real (correlated) symbols and simulated (independent) ones.

\subsection{A note on the estimation of the source probability distribution}
So far, we have considered that the probability $p$ of the source generating i.i.d. symbols is known to the encoder. In reality, this could not be the case and a measure of error is needed if the probability $r=\{r_1,\dots,r_N\}$ is used to encode symbols generated by the probability $p$. In the classical case, this is a well known problem. Assuming that it is possible to achieve the best encoding length which minimize the average length $L(0)$, i.e. $\ell_D(i) = -\log p_i$, then if the probability $r$ is practically used to encode symbols generated according to $p$, the average codewords length is simply given by \begin{align}
    H_1[p||r]=-\sum_{i=1}^N p_i \log_D r_i.
\end{align}
$H_1[p||r]$ is called cross-entropy. From this, it is possible to define the number of bits that are wasted by encoding according to $r$ as the difference between the cross-entropy (i.e. the actual average length) and the Shannon entropy (i.e. the lowest possible average length), thus getting the Kullback-Leibler divergence: \begin{align}
    D_{KL}[p||r]=H_1[p||r]-H_1[p]= \sum_{i=1}^N p_i \log \frac{p_i}{r_i}.
\end{align}
Following the same path, we would like to provide a measure in the case of an exponential average. While R\'enyi himself defined a generalized $D_{KL}$~\cite{renyi1961measures}, further analyzed in~\cite{van2014renyi} and~\cite{rached2001renyi}, and different definitions of a generalized cross-entropy exist~\cite{thierrin2022renyi}, we would like to define such quantities in the framework of data compression. In particular, the exponential average codeword length when $r$ is used to perform the compression is given by: \begin{align}
    \begin{split}
    H_q[p||r]&=\frac{1}{t}\log_D \sum_{i=1}^N p_i D^{-t\log_D (r^{(q)}_i)}\\
    &= \frac{q}{1-q}\log_D \sum_{i=1}^N p_i r_i^{q-1} + (1-q)H_q[r],
    \end{split}
\end{align}
where $r^{(q)}$ is the escort distribution of $r$, $q=1/(1+t)$ and $H_q[r]$ is the R\'enyi entropy of the distribution $r$. From this definition, it is possible to write a function for the error of encoding with distribution $r$ instead of the true $p$, as the difference between the actual exponential average length $H_q[p||r]$, and the lowest possible exponential average length $H_q[p]$, that would be obtained by the exact guessing of $p$, i.e. with $r=p$: 
\begin{align}
    \begin{split}
        \mathrm{ER}_q[p||r]&= H_q[p||r]-H_q[p]\\
        &=\frac{q}{1-q}\log_D \sum_{i=1}^N p_i r_i^{q-1} + (1-q)H_q[r]-H_q[p].
    \end{split}
\end{align}
It is easy to see that $\mathrm{ER}_q[p||p]=0$ $\forall q>0$ and that $\lim_{q\to 1}\mathrm{ER}_q[p||r]=D_{KL}[p||r].$
\begin{figure}[ht]
    \centering
    \includegraphics[scale=0.5]{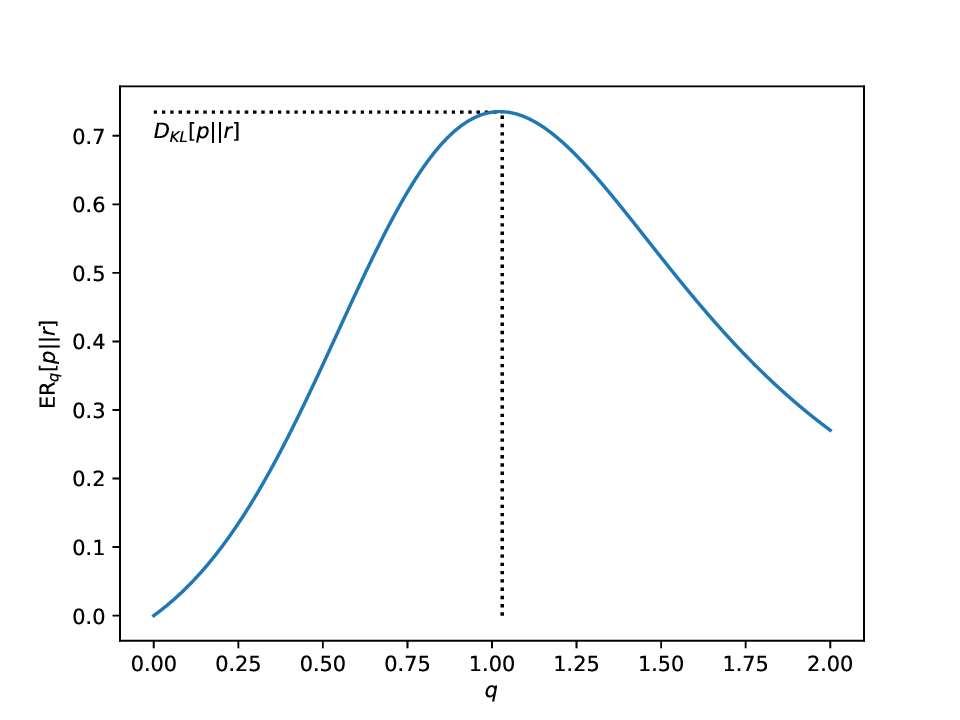}
    \caption{Error $\mathrm{ER}_q[p||r]$ for two instances of $p$ and $r$: $p_i\propto i^{-1}$ and $r_i\propto i^{-2}$. The horizontal dashed black line corresponds to $D_{KL}[p||r]$, which corresponds to $\mathrm{ER}_1[p||r].$}
    \label{fig:er}
\end{figure}
Figure~\ref{fig:er} shows the error function for varying $q$, with given $p$ and $r$ such that $p_i\propto i^{-1}$ and $r_i\propto i^{-2}$.

To our knowledge, despite the different definitions of generalized divergences and cross-entropies in the literature, the quantity $\mathrm{ER}_q$ has not been defined. Yet, it has a direct interpretation and provides a measure of how a wrong estimate of the probability $p$ propagates on the exponential average codeword length $L(t)$.

\section{Conclusions}
In this article, we have provided an operational scheme to encode sequences of symbols in order to minimize the exponential average codeword length. Our algorithm leads to an exponential average length per symbol that is arbitrarily close to the R\'enyi entropy of the source distribution. While our theoretical analysis relies on the symbols being i.i.d., we have shown that it provides advantageous results even in the case of correlated symbols, with respect to the usual $q=1$ Shannonian compressor. Moreover, we have detailed a possible application of the exponential average, based on its connection with the cumulant generating function of the source's probability distribution. Namely, if the encoder's priority is to minimize the risk of exceeding a certain codewords' threshold length, minimizing the exponential average is a better solution than minimizing the linear average. Even if all our theoretical considerations are based on the hypothesis that the symbols are i.i.d. distributed and that the encoder knows the true source distribution $p$,  we have both shown empirically that AC$_q$ is advantageous also in the presence of correlations and provided a measure of the error when the encoder guesses the incorrect source distribution. However, a theoretical description explaining quantitatively how correlations lead to an optimal $q$ different from $q_t$, and what is the expected error of guessing the source distribution given a certain (``small'') training dataset still lacks and can be the object of future studies.

\section*{Acknowledgements}
The work of P.F. and D.G. has been supported by the European Union -- Horizon 2020 Program
under the scheme ``INFRAIA-01-2018-2019 -- Integrating Activities for
Advanced Communities'', Grant Agreement n. 871042, ``SoBigData++: European
Integrated Infrastructure for Social Mining and Big Data Analytics'' \url{http://www.sobigdata.eu}, by the NextGenerationEU -- National Recovery and Resilience
Plan (Piano Nazionale di Ripresa e Resilienza, PNRR) -- Project: ``SoBigData.it -- Strengthening the Italian RI for Social Mining and Big Data Analytics''
-- Prot. IR0000013 -- Avviso n. 3264 del 28/12/2021. P.F. also acknowledges support by the spoke ``FutureHPC
\& BigData'' of the ICSC – Centro Nazionale di Ricerca in High-Performance
Computing, Big Data and Quantum Computing funded by European Union –
NextGenerationEU – PNRR. A.S. and D.G. acknowledge support from the Dutch Econophysics Foundation (Stichting Econophysics, Leiden, the Netherlands).

\bibliographystyle{unsrt}
\bibliography{main}

\begin{thebibliography}{10}

\bibitem{shannon1948mathematical}
Claude~Elwood Shannon.
\newblock A mathematical theory of communication.
\newblock {\em The Bell system technical journal}, 27(3):379--423, 1948.

\bibitem{kolmogorov1930notion}
Andre{\u\i}~Nikolaevich Kolmogorov and Guido Castelnuovo.
\newblock {\em Sur la notion de la moyenne}.
\newblock G. Bardi, tip. della R. Accad. dei Lincei, 1930.

\bibitem{nagumo1930klasse}
Mitio Nagumo.
\newblock {\"U}ber eine klasse der mittelwerte.
\newblock In {\em Japanese journal of mathematics: transactions and abstracts}, volume~7, pages 71--79. The Mathematical Society of Japan, 1930.

\bibitem{aczel1975measures}
Joseph Acz{\'e}l and Zolt{\'a}n Dar{\'o}czy.
\newblock {\em On measures of information and their characterizations.}
\newblock Academic Press, New York, 1975.

\bibitem{campbell1966definition}
LL~Campbell.
\newblock Definition of entropy by means of a coding problem.
\newblock {\em Zeitschrift f{\"u}r Wahrscheinlichkeitstheorie und verwandte Gebiete}, 6(2):113--118, 1966.

\bibitem{hardy1952inequalities}
Godfrey~Harold Hardy, John~Edensor Littlewood, and George P{\'o}lya.
\newblock {\em Inequalities}.
\newblock Cambridge university press, 1952.

\bibitem{Morales_2023}
Pablo~A Morales, Jan Korbel, and Fernando~E Rosas.
\newblock Thermodynamics of exponential kolmogorov–nagumo averages.
\newblock {\em New Journal of Physics}, 25(7):073011, jul 2023.

\bibitem{meiser2022synthetic}
Linda~C Meiser, Bichlien~H Nguyen, Yuan-Jyue Chen, Jeff Nivala, Karin Strauss, Luis Ceze, and Robert~N Grass.
\newblock Synthetic dna applications in information technology.
\newblock {\em Nature communications}, 13(1):352, 2022.

\bibitem{mishra2020compressed}
Pooja Mishra, Chiranjeev Bhaya, Arup~Kumar Pal, and Abhay~Kumar Singh.
\newblock Compressed dna coding using minimum variance huffman tree.
\newblock {\em IEEE Communications Letters}, 24(8):1602--1606, 2020.

\bibitem{buffer1}
Frederick Jelinek.
\newblock Buffer overflow in variable length coding of fixed rate sources.
\newblock {\em IEEE Transactions on Information Theory}, 14(3):490--501, 1968.

\bibitem{buffer2}
Pierre Humblet.
\newblock Generalization of huffman coding to minimize the probability of buffer overflow (corresp.).
\newblock {\em IEEE Transactions on Information Theory}, 27(2):230--232, 1981.

\bibitem{iridium}
Liam David and Anonto Zaman.
\newblock Simulating iridium satellite coverage for cubesats in low earth orbit.
\newblock {\em in Proc. 32nd Annu. AIAA/USU Conf. Small Satellites}, 2018.

\bibitem{baer2008optimal}
Michael~B Baer.
\newblock Optimal prefix codes for infinite alphabets with nonlinear costs.
\newblock {\em IEEE Transactions on Information Theory}, 54(3):1273--1286, 2008.

\bibitem{campbell1965coding}
L~Lorne Campbell.
\newblock A coding theorem and r{\'e}nyi's entropy.
\newblock {\em Information and control}, 8(4):423--429, 1965.

\bibitem{tsallis2009introduction}
Constantino Tsallis.
\newblock {\em Introduction to nonextensive statistical mechanics: approaching a complex world}.
\newblock Springer, 2009.

\bibitem{beck_schögl_1993}
Christian Beck and Friedrich Schögl.
\newblock {\em Thermodynamics of Chaotic Systems: An Introduction}.
\newblock Cambridge Nonlinear Science Series. Cambridge University Press, 1993.

\bibitem{somazzi2023learn}
Andrea Somazzi and Diego Garlaschelli.
\newblock Learn your entropy from informative data: an axiom ensuring the consistent identification of generalized entropies.
\newblock {\em arXiv preprint arXiv:2301.05660}, 2023.

\bibitem{khincin}
A.I Khinchin.
\newblock Mathematical foundation of information theory.
\newblock {\em Dover Publications, New York}, 1957.

\bibitem{jizba2020shannon}
Petr Jizba and Jan Korbel.
\newblock When shannon and khinchin meet shore and johnson: Equivalence of information theory and statistical inference axiomatics.
\newblock {\em Physical Review E}, 101(4):042126, 2020.

\bibitem{merhav1991universal}
Neri Merhav.
\newblock Universal coding with minimum probability of codeword length overflow.
\newblock {\em IEEE Transactions on Information Theory}, 37(3):556--563, 1991.

\bibitem{blumer1988renyi}
Anselm~C Blumer and Robert~J McEliece.
\newblock The r{\'e}nyi redundancy of generalized huffman codes.
\newblock {\em IEEE Transactions on Information Theory}, 34(5):1242--1249, 1988.

\bibitem{bercher2009source}
J-F Bercher.
\newblock Source coding with escort distributions and r{\'e}nyi entropy bounds.
\newblock {\em Physics Letters A}, 373(36):3235--3238, 2009.

\bibitem{moffat2020large}
Alistair Moffat and Matthias Petri.
\newblock Large-alphabet semi-static entropy coding via asymmetric numeral systems.
\newblock {\em ACM Transactions on Information Systems (TOIS)}, 38(4):1--33, 2020.

\bibitem{witten1987arithmetic}
Ian~H Witten, Radford~M Neal, and John~G Cleary.
\newblock Arithmetic coding for data compression.
\newblock {\em Communications of the ACM}, 30(6):520--540, 1987.

\bibitem{ferragina_2023}
Paolo Ferragina.
\newblock {\em Pearls of Algorithm Engineering}.
\newblock Cambridge University Press, 2023.

\bibitem{cover1999elements}
Thomas~M Cover.
\newblock {\em Elements of information theory}.
\newblock John Wiley \& Sons, 1999.

\bibitem{renyi1961measures}
Alfr{\'e}d R{\'e}nyi.
\newblock On measures of entropy and information.
\newblock In {\em Proceedings of the Fourth Berkeley Symposium on Mathematical Statistics and Probability, Volume 1: Contributions to the Theory of Statistics}, volume~4, pages 547--562. University of California Press, 1961.

\bibitem{van2014renyi}
Tim Van~Erven and Peter Harremos.
\newblock R{\'e}nyi divergence and kullback-leibler divergence.
\newblock {\em IEEE Transactions on Information Theory}, 60(7):3797--3820, 2014.

\bibitem{rached2001renyi}
Ziad Rached, Fady Alajaji, and L~Lorne Campbell.
\newblock R{\'e}nyi's divergence and entropy rates for finite alphabet markov sources.
\newblock {\em IEEE Transactions on Information theory}, 47(4):1553--1561, 2001.

\bibitem{thierrin2022renyi}
Ferenc~Cole Thierrin, Fady Alajaji, and Tam{\'a}s Linder.
\newblock R{\'e}nyi cross-entropy measures for common distributions and processes with memory.
\newblock {\em Entropy}, 24(10):1417, 2022.

\end{thebibliography}

\end{document}